\documentclass [11pt]{article}
\usepackage{amsmath,amsthm,amsfonts,amscd,eucal,latexsym,amssymb}
\usepackage{epsfig}   
\input xy
\xyoption{all}
\def\sp{\hskip -5pt} 
\def\spa{\hskip -3pt}

\newsymbol\bt 1202           

\def\cF{{\ca F}}

\def\cH{{\ca H}}
\def\cD{{\ca D}}

\def\cS{{\ca S}}

\def\cF{{\ca F}}

\def\cW{{\ca W}}

\def\sS{{\mathsf S}}

\def\bC{{\mathbb C}}           

\def\bR{{\mathbb R}}

\def\bS{{\mathbb S}}

\newsymbol \rest 1316         
 
\def\gB{{\mathfrak B}}

\def\gH{{\mathfrak H}}

\def\beq{\begin{eqnarray}}
\def\eeq{\end{eqnarray}}

\newcommand{\ca}[1]{{\cal #1}}         

\def\z{\zeta}
\def\bz{\overline{\zeta}}

\def\scri{\Im^+}         
\def\tg{\tilde{g}}
\def\tM{\tilde{M}}

\def\lie{\pounds}

\newcounter{proposition}[section]
\newcounter{theorem}[section]
\newcounter{lemma}[section]
\newcounter{definition}[section]
\newcounter{remark}[section]

\def\theproposition{\thesection.\arabic{proposition}}
\def\thetheorem{\thesection.\arabic{theorem}}
\def\thelemma{\thesection.\arabic{lemma}}
\def\thedefinition{\thesection.\arabic{definition}}
\def\theremark{\thesection.\arabic{remark}}

\def\s #1 {\section{#1}}

\def\ssa #1 {\ifhmode{\par}\fi\refstepcounter{subsection}
  \noindent {\bf\thesubsection}. {\em #1}.\quad
  \addcontentsline{toc}{subsection}{\protect\numberline{\thesubsection} #1}%
  }

\def\ssb #1 {\ifhmode{\par}\fi\refstepcounter{subsection}
  \noindent {\bf\thesubsection.} {\em #1.}\quad
  \addcontentsline{toc}{subsection}{\protect\numberline{\thesubsection} #1}%
  }

\def\proposizione {\ifhmode{\par}\fi\refstepcounter{proposition}
  \noindent {\bf Proposition \theproposition}. \quad}
\def\teorema {\ifhmode{\par}\fi\refstepcounter{theorem}
  \noindent {\bf Theorem \thetheorem}. \quad}
\def\lemma {\ifhmode{\par}\fi\refstepcounter{lemma}
  \noindent {\bf Lemma \thelemma}. \quad}
\def\definizione {\ifhmode{\par}\fi\refstepcounter{definition}
  \noindent {\bf Definition \thedefinition}. \quad}
\def\remark {\ifhmode{\par}\fi\refstepcounter{remark}
  \noindent {\bf Remark \theremark}. \quad}

\begin{document} 
 
\hfill{\sl November 2006, Preprint  UTM-706} \\

 
\par 
\LARGE 
\noindent 
{\bf Some recent results in linear scalar quantum field
theory in globally hyperbolic asymptotically
flat spacetimes} \\
\par 
\normalsize 
 
 

\noindent 
{\bf Valter Moretti} \\
\par
\small
\noindent Dipartimento di Matematica, Universit\`a di Trento, 
 and Istituto Nazionale di Alta Matematica ``F.Severi''  unit\`a locale  di Trento,
 and  Istituto Nazionale di Fisica Nucleare  Gruppo Collegato di Trento,\\ via Sommarive 14  
I-38050 Povo (TN), Italy. \\
E-mail:  moretti@science.unitn.it\\ 
 \normalsize

\small 
\noindent {\bf Abstract}. {
\noindent The content of this paper is that of an invited plenary talk at the  
{\em XVII SIGRAV Conference held in Torino, September 4-7, 2006}.\\
Some recent results obtained by the author and 
collaborators about QFT in asymptotically flat spacetimes at null  infinity are reviewed.
In particular it is shown that bosonic QFT can be defined on the null boundary $\scri$ of any asymptotically 
flat spacetime $M$. This theory admits a state $\lambda$ which is uniquely determined from invariance under
BMS group and a BMS-energy positivity requirement. There is a nice interplay with bosonic 
(massless, conformally coupled)
QFT defined in the bulk spacetime. In particular, under suitable further requirements,  the universal state $\lambda$ induces 
in the bulk spacetime $M$ a state,
$\lambda_M$, which enjoys the following remarkable properties.
It reduces to standard Minkowski vacuum whenever $M$ is Minkowski spacetime and
in the general case, it is invariant under the group of isometries 
of the spacetime $M$, it is a ground state (i.e. it satisfies the positive energy condition) with respect to any timelike Killing time of $M$
without zero-modes, finally $\lambda_M$ enjoys the global Hadamard property so that it is suitable for locally covariant 
perturbative renormalization procedures.\\}

\section{Asymptotically Flat Spacetimes and $BMS$ group}
Asymptotically flat spacetimes -- by definition -- have a certain {\em asymptotic structure}.
The main motivation of the works in the references \cite{DMP,CMP5,last} has been investigating if that asymptotic structure
{\em determines canonically preferred states in every asymptotically flat spacetime} for (linear) 
scalar quantum field theory.  The properties of those states have been also focused.
Let us remind the main definition \cite{As80} (See \cite{Wald1} for further discussions and references).\\

\noindent {\bf Definition 1.} \label{sefasympt}  ({\bf Asymptotically flat vacuum spacetime at future null infinity}) 
{\em A four dimensional spacetime $(M,g)$ is called {\bf asymptotically flat vacuum spacetime at future null infinity} if:\\
{\bf (a)} $(M,g)$ can be seen as an embedded submanifold of a larger spacetime $(\tilde{M},\tilde{g})$ 
with $\tilde{g}\spa\rest_M= \Omega^2  g$, $\Omega$ being a smooth function on $\tM$, strictly positive on $M$. \\
{\bf (b)} $\scri:= \partial M$, called {\bf future null infinity} of $M$, is a $3$-dim $\tilde {M}$-submanifold  satisfying 
    
     (i) $\Omega\spa\rest_{\scri} =0$ but $d\Omega\spa\rest_{\scri} \neq 0$,
    
     (ii) $\scri \cap \tilde{J}^-(M) = \emptyset$,
    
      (iii) $\scri$ diffeomorphic to $\bS^2\times \bR$, and it is 
union of  the {\bf null} curves tangent to $n := \nabla \Omega$.
These curves are {\bf complete null geodesics} for a certain choice of $\Omega$.\\
{\bf (c)} About $\scri$, $(M,g)$ is strongly causal and  satisfies $Ric(g) =0$.}\\

\noindent In fact $\scri$ is a $3$-dim. null submanifold of $\tilde{M}$  
with {\em degenerate metric} $\tilde{h}$ induced by $\tilde{g}$.
$\scri$ is the {\em conic surface} indicated by $\:\mbox{I}^+\:$ in the figure. 
The tip is {\em not} a point of $\tM$.
\begin{figure}[th]
\begin{center}
\includegraphics[scale=.7]{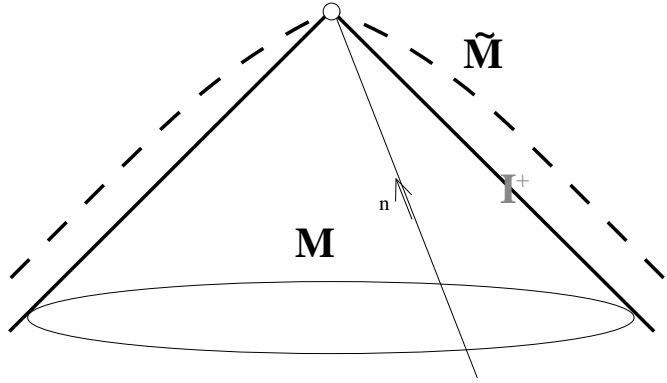}
\end{center}
\caption{Asimptotically flat spacetime.}
\end{figure}


\noindent  There are {\em gauge transformations} permitted by the definition:  $\Omega \to \omega \Omega$  with  
$\omega>0$ in a neighborhood of $\scri$, 
$$
\scri \to \scri \:,\:\:\:\:\: \tilde{h} \to \omega^2 \tilde{h} \:,\:\:\:\:\: n \to \omega^{-1} n \:.
$$
For a fixed  a.f. spacetime $(M,g)$, $C_{(M,g)}$ is the class of all triples $\{(\scri,\tilde{h}, n)\}$
connected by gauge transformations. $C_{(M,g)}$ encodes the whole geometric extent 
of $\scri$.\\
 It is important to stress that there is no physically preferred element of $C_{(M,g)}$ therefore
  it is mathematical convenient to pass from
$\Omega$ to $\Omega_B = \omega_B \Omega$ such that $(\scri,\tilde{h}_B,n_B)$ reads
\begin{center}$\displaystyle \scri = \bR \times \bS^2\:, \quad \tilde{h}_B =  d\theta \otimes d\theta +\sin^2\theta d\varphi \otimes d\varphi\:, \quad n_B = \frac{\partial \:\:}{\partial u}$\end{center}
in coordinates $u\in \bR$ (futuredirected), $(\theta,\varphi) \in \bS^2$. $(u,\theta,\varphi)$ is a 
{\bf Bondi frame} on $\scri$.\\
Notice that if  $(M_1,g_1) \neq (M_2,g_2)$, for a suitable choice of $(\scri_1,\tilde{h}_1, n_1)\in C_{(M_1,g_1)}$
and $(\scri_2,\tilde{h}_2, n_2)\in C_{(M_2,g_2)}$,
$\exists$ diffeomorphism $\gamma: \scri_1 \to \scri_2$ with 
 \begin{center}$\gamma(\scri_1) = \scri_2 \:,\:\:\:\:\: \gamma^* \tilde{h}_1=\tilde{h}_2 \:,\:\:\:\:\:\gamma^* 
n_1=n_2$ \end{center} 
In this sense the class $C=C_{(M,g)}$ is {\bf universal} for all a.f. spacetimes.\\
 We have the following subsequent definition \cite{BMS70-80}:\\

\noindent{\bf Definition 2.}  ({\bf Bondi-Metzner-Sachs (BMS) group}.)
 {\em $G_{BMS}$ is the group of diffeomorphisms $\chi : \scri \to \scri$ preserving the universal structure of $\scri$, i.e. {\bf $\chi$ are isometries up to gauge transformations}:
 $$\chi (\scri) = \scri\:, \quad  \chi^*\tilde{h} = \omega^2_\chi \tilde{h}\:,\quad  \chi^*n = \omega^{-1}_\chi n
\qquad  \mbox{for some $\omega_\chi >0$.}$$}
 
\noindent Let us examine the  structure of $G_{BMS}$. Fix a {\em Bondi frame} $(u,\varphi,\theta)$ on $\scri$
and pass to complex coordinates on the sphere $\bS^2$
 $(u,\varphi,\theta) \leftrightarrow (u,\z,\bz)$,  with 
 $\z := e^{i\varphi}\cot(\theta/2) \in \bC \cup \{+\infty\}\:,$ the usual Riemann sphere. 
In this framework $G_{BMS}$ has the structure of a semi-direct product, 
$G_{BMS} = SO(3,1)\sp\uparrow  \times C^\infty(\bS^2)$, where the group composition law is defined in this way: 
\begin{eqnarray}
G_{BMS} &\ni& (\Lambda, f) : (u,\z,\bz) \mapsto (u',\z',\bz')\:, \nonumber \\
 u' &:=& K_\Lambda(\z,\bz)(u + f(\z,\bz))\:,\quad
\z' \:\::= \:\: \frac{a_\Lambda\z + b_\Lambda}{c_\Lambda\z +d_\Lambda}\:,\nonumber \\
 K_\Lambda(\z,\bz) &:=&  \frac{(1+\z\bz)}{|a_\Lambda\z + b_\Lambda|^2 +|c_\Lambda\z +d_\Lambda|^2}
\label{K}\:\: \: \mbox{and}\:\:\:\:
\Pi \left[
\begin{array}{cc}
  a_\Lambda & b_\Lambda\\
  c_\Lambda & d_\Lambda 
\end{array}
\right]  = \Lambda \:.\nonumber
\end{eqnarray}
Above $SO(3,1)\sp\uparrow$ is the special orthochronous Poincar\'e group and $\Pi : SL(2,\bC) \to SO(3,1)\sp\uparrow$ is the standard covering homomorphism.\\
 It is worth stressing that pairs of Bondi frames are connected by transformations of a subgroup of $G_{BMS}$.
 Moreover notice that the  decomposition $G_{BMS} = SO(3,1)\sp\uparrow  \times C^\infty(\bS^2)$ depends on the used frame,
however the subgroup:
$$T^4 := \mbox{{real span of the first} {\bf $4$} {\bf spheric harmonics}} \subset C^\infty(\bS^2)$$
 is a {\bf normal} $G_{BMS}$-subgroup. It is called the subgroup of {\bf $4$-translations}.
 Thus $T^4$ is $BMS$-{\bf invariant}, i.e. {\bf independent} from used reference (Bondi) frames, and so it may 
have physical interest. Another important fact is that there is a {\em Minkowskian decomposition}
$T^4 \ni \alpha = \displaystyle \sum_{\mu=0}^3 \alpha^\mu Y_\mu$ 
where $\{Y_\mu\}$ is a certain real basis of  $T^4$. In this framework, if $\alpha,\alpha' \in T^4$
the {\em Minkowskian scalar product} $(\alpha,\alpha') := \sum_{\mu,\nu}\eta_{\mu\nu} \alpha^\mu {\alpha'}^\nu$ turns out to
be
$BMS$-{\bf invariant},
i.e. invariant under $\alpha \mapsto  := g\circ \alpha \circ g^{-1}$, $g\in
 G_{BMS}$ and $\alpha \in C^\infty(\bS^2)$.  ($\eta = diag (-1,1,1,1)$.)\\
We conclude that there is a $BMS$-{\bf invariant} decomposition of $T^4$ and thus  we have {\em spacelike, (fut./past. directed) timelike, 
(fut./past. directed) null} $4$-translations. Time orientation is induced by that of $(M,g)$ (see \cite{last} for details.)\\

\noindent {\bf Remarks}.\\
{\bf (1)} A timelike future-directed {\bf direction} $\{\lambda \alpha\}_{\lambda\in \bR}$ in $T^4$ 
individuates a Bondi frame $(u,\z,\bz)$ uniquely: that where the action of $\{\lambda \alpha\}_{\lambda\in \bR}$ 
is $u\mapsto u+\lambda$.\\
{\bf (2)} The orthochronous proper Poincar\'e group ${\cal P} :=SO(3,1)\sp\uparrow  \times T^4$ is {\it not} 
a normal subgroup of $G_{BMS}$, hence there is {\bf no} physical way 
to pick out a preferred Poincar\'e subgroup of $G_{BMS}$. \\
{\bf (3)} It is known that $G_{BMS}$ {\em encodes  the bulk symmetries of each fixed asymptotically 
flat vacuum at null infinity spacetime $(M,g)$} as well as all the
{\em asymptotic symmetries} of all bulk spacetimes. Indeed
the following result holds true \cite{GerochAshtekarXanthopoulos80}.\\

\noindent {\bf Theorem 1.} 
{\it Let $(M,g)$ be asymptotically flat vacuum at null infinity spacetime. The following holds.\\
{\bf (a)} If $\xi$ is a Killing vector field of the bulk $M$, then $\xi$ 
smoothly extends to a  vector field $\tilde{\xi}$, different from the zero vector field and
tangent to $\scri$,
which generates a one-parameter subgroup of $G_{BMS}$.\\
{\bf (b)} In that way, the  isometry group of $M$ is mapped {\bf into} a subgroup $G_M\subset G_{BMS}$ with:\\
\indent (i) $G_M$ is isomorphic to a subgroup of a certain Poincar\'e group 
 ${\cal P}\subset G_{BMS}$ where ${\cal P}$ generally
  {\em depends} on the particular spacetime $M$,\\
  \indent (ii) only proper $4$-translations are admitted in $G_M \cap C^\infty(\bS^2)$.\\ 
 {\bf (c)} $\tilde{\xi}$ generates an one-parameter subgroup of $G_{BMS}$, if and only if it smoothly extends back to a
 field $\xi$ in $M$ with $\Omega^2 \lie_\xi g \to 0$ smoothly approaching $\scri$.}\\

\noindent 
The item (c) states, in fact, that $G_{BMS}$ contains the so called so-called {\bf asymptotic symmetries} of the physical
spacetime $M$ \cite{Wald1}.
The BMS group is the group of invariance of any physical theory defined on $\scri$: mathematical objects 
defined on $\scri$ may have physical sense only if they are invariant under $G_{BMS}$.

\section{Weyl quantization on $\scri$ and interplay with QFT in the bulk: the state $\lambda_M$.}
In \cite{DMP} it has been established that it is possible to define a bosonic QFT for a field defined on $\scri$.
The approach is that algebraic based on Weyl quantization \cite{Wald}. The ingredients are the following ones.
A {\em real symplectic space}: $(\sS(\scri),\sigma)$, where  $\sigma$ nondegenerate {\em symplectic form}, 
whereas the real vector space $\cS(\scri)$ is defined as (for a fixed Bondi frame):
\begin{eqnarray} \sS(\scri) := \left\{ \left.\psi \in C^\infty(\scri)\: \:\right|\:\: \psi\:, \partial_u 
\psi \in  L^2(\bR\times \bS^2, du \wedge \epsilon_{\bS^2}(\z,\bz)) \right\}, \label{S}\end{eqnarray} 
whereas the explicit form of $\sigma$ is the following:
\begin{eqnarray}  \displaystyle \sigma(\psi_1,\psi_2) := \int_{\bR\times \bS^2} 
\left(\psi_2 \frac{\partial\psi_1}{\partial u}  - 
\psi_1 \frac{\partial\psi_2}{\partial u}\right) 
du \wedge \epsilon_{\bS^2}(\z,\bz)\:, \label{sigma} \end{eqnarray} 
where $\epsilon_{\bS^2}$ is the standard measure on a unit $2$-sphere.
There is a  {\bf weighted representation of $G_{BMS}$}, $A_g : C^\infty(\scri) \to C^\infty(\scri)$, 
$g\in G_{BMS}$, acting on the considered symplectic space:
\begin{eqnarray}  (A_g\psi)(u,\z,\bz):=(K_\Lambda^{-1}\cdot  \psi)(g^{-1}(u,\z,\bz))\:\:\: \mbox{(notice the weight 
$K_\Lambda^{-1}$, $g= (\Lambda,f))$}\:. \label{A}\end{eqnarray} 
Notice that  $A_g\spa\rest_{\sS(\scri)} \subset \sS(\scri)$, moreover, 
due to the weight $K_\Lambda^{-1}$, the $G_{BMS}$ representation $A$ {\em preserves} the symplectic form of $\sigma$,
so that, in particular, the mentioned structure does not depend on the used Bondi frame.

With those ingredients one defines the (unique)
{\bf Weyl $C^*$-algebra} $\cW(\sS(\scri),\sigma)$ with generators $W(\psi)\neq 0$, $\psi \in \sS(\scri)$, satisfying 
Weyl relations (also known as {\bf CCR}):
$$ W(-\psi)= W(\psi)^*\:,\quad\quad W(\psi)W(\psi') = e^{i\sigma(\psi,\psi')/2} W(\psi+\psi')\:.
$$
Moreover the representation  $A$ induces a {\em $*$-automorphism $G_{BMS}$-representation} 
\begin{eqnarray} \alpha : \cW(\sS(\scri),\sigma) \to \cW(\sS(\scri),\sigma)\label{alpha} \end{eqnarray} 
uniquely individuated by the requirement
$\alpha_g(W(\psi)):= W(A_{g^{-1}}\psi)$.\\
In order to find a possible physical meaning of the theory constructed above,
 a natural question arises: Spacetime physics is BMS-invariant so,
{\em are there  BMS-invariant (quasifree) algebraic states on $\cW(\sS(\scri),\sigma)$?}\\
An answer has been found in \cite{DMP}. There it has demonstrated that, in fact, there is a (quasifree pure) 
$BMS$-invariant state. Let us summarize this result. 
Consider the  quasifree pure state $\lambda$ on $\cW(\sS(\scri),\sigma)$ uniquely induced by linearity and continuity from:
\begin{eqnarray} \lambda(W(\psi)) = e^{-\mu_\lambda(\psi,\psi)/2}\:,\quad \mu_\lambda(\psi_1,\psi_2):=  -i \sigma(\overline{\psi_{1+}},\psi_{2+})\:,
\quad\psi \in \sS(\scri)\:, \label{lambda} \end{eqnarray}
where $\psi_+$  is the {\em positive $u$-frequency part} of $\psi$, with respect to any (arbitrarily fixed) Bondi frame
defined on $\scri$. The positive frequency part is obtained performing the usual Fourier transform with respect to the 
variable $u$ considered as a ``time'' coordinate (see \cite{DMP} for details).
Then we pass to focus on the GNS representation $(\gH_\lambda, \Pi_\lambda, \Upsilon_\lambda)$. It turns out that the Hilbert
space 
  $\gH_\lambda$ is a bosonic Fock space $\cF_+(\cH)$ with $1$-particle Hilbert space 
$\cH \equiv L^2(\bR^+\times \bS^2; dE \otimes \epsilon_{\bS^2})\quad$ (containing $u$-Fourier transforms $\widehat{\psi}_+ $),
the GNS Cyclic vector is the Fock vacuum $\Upsilon_\lambda$. 
 Since the GNS representation is a Fock representation, $\lambda$ is a regular state and
 {\em symplectically-smeared} field operators  $\Psi(\psi)$ with 
 $\Pi_\lambda(W(\psi)) = e^{-i\overline{\Psi(\psi)}}$ can be defined using Stone theorem directly (see e.g. \cite{Wald} for the 
 general theory).\\
 In this context we have the following theorem established in \cite{DMP}.\\

\noindent {\bf Theorem 2.}
{\it Referring to the Weyl algebra  $\cW(\sS(\scri),\sigma)$, 
its GNS representation $(\gH_\lambda, \Pi_\lambda, \Upsilon_\lambda)$, the BMS representation $\alpha$ (\ref{alpha}),
and the state $\lambda$ (\ref{lambda}), the following facts are valid.\\
 {\bf (\bf a)} $\lambda$ is {\bf $G_{BMS}$-invariant}: $\lambda(\alpha_g(a)) = \lambda(a)$ if 
$g\in G_{BMS}$ and $a\in \cW(\sS(\scri),\sigma)$.\\
{\bf (\bf b)} The unique unitary representation of $G_{BMS}$
leaving $\Upsilon_{\lambda}$ invariant (that is $U_g \Upsilon_\lambda = \Upsilon_\lambda$) and implementing $\alpha$ 
(i.e. $U_g \Pi_\lambda(a) U^*_g = \Pi_\lambda(\alpha_g(a))$) is that induced by: 
\begin{eqnarray}    \left(U_{(\Lambda,f)}\widehat{\psi}_+\right)(E,\z,\bz) = 
\frac{e^{iE K_{\Lambda}(\Lambda^{-1}(\z,\bz))f(\Lambda^{-1}(\z,\bz))}}{
  \sqrt{K_{\Lambda}(\Lambda^{-1}(\z,\bz))}}
 \widehat{\psi}_+\left(E K_{\Lambda}\left(\Lambda^{-1}(\z,\bz)\right),\Lambda^{-1}(\z,\bz)\right) \:. 
 \label{U}\end{eqnarray} 
 $\widehat{\psi}_+$ being the positive-frequency part of $\psi\in \sS(\scri)$ in Fourier representation ($E$ being 
 the conjugate variable with $u$).\\
{\bf (\bf c)} Making $G_{BMS}$ topological  
 equipping  $C^\infty(\bS^{2}) \subset G_{BMS}$ with test-function Fr\'echet topology, consider the representation  
 $G_{BMS} \ni g \mapsto U_g$, then one has:

 (i) it is {\em irreducible and strongly continuous},
 
 (ii) it is a {\em Wigner-Mackey-like  representation} associated with a scalar representation of the 
 little group, $\Delta \subset SL(2,\bC)$, the double covering of the $2D$
 Euclidean group,
 
(iii) it is defined on an orbit in space of characters $\chi$ with $m^2_{BMS}(\chi) = 0$. }\\

\noindent To make a comment to (c) we stress that
  the Abelian $G_{BMS}$-subgroup $C^\infty(\bS^2)$ is {\it infinite dimensional and  non-locally-compact},
 but
Mackey machinery {\it works anyway} as proved in \cite{McCarthyPRSL}.
Moreover, concerning (iii)  notice that the characters, $\chi_\beta$, 
are labeled by distributions (here $\cD'(\bS^2)$ is the dual space of $C_0^\infty(\bS^2)$)
$\beta \in \cD'(\bS^2)$, $\chi_\beta(\alpha) = e^{i\beta(\alpha)}$, 
 $\forall \alpha \in C^\infty(\bS^2)$. Therefore, in the space of characters can be defined a {\em BMS-invariant mass}: 
 $m^2_{BMS}(\beta) :=  -\eta^{\mu\nu}\beta(Y_\mu)\beta(Y_\nu)$ which turns out to be invariant with respect to
  the dual action $(g(\beta))(\alpha)  := \beta(g(\alpha))$, $\forall g\in G_{BMS}$. This is a notion of mass
  which is {\em a priori} independent from that invariant under the action of Poincar\'e group.

Now a question arises naturally: {\em Is there any relation with massless particles propagating 
in the bulk spacetime?}
The answer is positive as established in \cite{DMP}:  the fields $\psi$ on $\scri$ are ``extensions'' 
of linear, massless, conformally coupled fields
in the bulk spacetime and the action of symmetries on the fields in the bulk is 
equivalent to the action of $G_{BMS}$ on the
associated fields on $\scri$.\\
 
\noindent {\bf Theorem 3.} 
{\it Let $(M,g)$ be an asymptotically flat vacuum at future null infinity spacetime  with associated unphysical spacetime 
$(\tM,\tg= \Omega^2 g)$. Assume that both $M,\tM$ are {\em globally hyperbolic}. Consider Weyl QFT in $(M,g)$ 
based on the symplectic space $(\cS(M),\sigma_M)$.  $\cS(M)$ is the space of real
smooth, compactly supported on Cauchy surfaces, solutions $\phi$ of massless, conformally-coupled, K-G equation\\
$$\displaystyle \Box \phi - \frac{1}{6} R \phi =0\quad \mbox{in $M$.}$$
with isometry-invariant symplectic form:\\
$$\sigma_M(\phi_1,\phi_2) := \int_\Sigma \left(\phi_2 \nabla_N \phi_1 - \phi_1 \nabla_N \phi_2\right)\: 
d\mu^{(S)}_g\:,$$ $\Sigma$ being any
Cauchy surface of $M$.  Then:\\
{\bf (a)} $\phi$ vanishes approaching $\scri$ but $\Omega_B^{-1}\phi$ extends to a smooth field $\psi := \Gamma\phi$ 
on $\scri$ uniquely, $\Gamma$ being linear.\\
{\bf (b)} If $\{g_t\}$ is a $1$-parameter group of $M$-isometries and $\{g'_t\}$ the associated $G_{BMS}$-subgroup:
the action of $\{g_t\}$ on $\phi$ ($\phi \mapsto \alpha_{g_t}(\phi) := \phi \circ g_{-t}$) is equivalent to the action of 
$A_{g'_t}$ on $\psi = \Gamma\phi$,
\begin{center} $\displaystyle A_{g'_t}(\psi) = \Gamma(\alpha_{g_t}(\phi))  
\quad \mbox{if $\quad\psi = \Gamma\phi$}$\:.\end{center}}

\noindent {\bf Remark}. The more usual point of view in considering QFT in globally hyperbolic spacetimes $(M,g)$
is that based on field operators $\Phi(f)$ smeared with smooth compactly supported functions $f\in C_0^\infty(M)$, instead of
with solutions of Klein-Gordon equation of $\cS(M)$. Actually the two points of view, for linear theories, are {\em completely equivalent},
 see \cite{Wald} for instance. If $\phi \in \sS(M)$, the  Weyl generators $W(\phi)$ are to be formally understood 
 as $W(\phi):= e^{-i\sigma_M(\Phi,\phi)}$, where $\sigma_M(\Phi,\psi)$ is the {\bf field operator symplectically smeared} with elements of $ \sS(M)$.
 In this context, it turns out that $\Phi(f):= \sigma_M(\Phi,Ef)$ where $E = A-R : C^\infty_0(M) \to \cS(M)$ is  the
{\bf causal propagator} (or ``advanced-minus retarded'' fundamental solution) of Klein-Gordon operator \cite{Wald}.\\

\noindent To go on, assume  that furthermore $\Gamma : \sS(M) \to \sS(\scri)$ is an {\em injective
symplectomorphism} i.e.
 (H1) $Range [\Gamma] \subset \sS(\scri)\qquad$
 (H2) $\sigma_\scri(\Gamma
\phi_1,\Gamma \phi_2) = \sigma_M (\phi_1,\phi_2)$.\\
In this case one finds that {\it the field observables of the bulk $M$ 
can be identified with observables of the boundary $\scri$}.
 More precisely \cite{DMP}: $\exists !$ a (isometric) $*$-homomorphism from the  Weyl algebra $\cW(\sS(M), \sigma_M)$ 
of field observables of the bulk, to the Weyl-algebra $\cW(\sS(\scri), \sigma)$: 
$$\imath_\Gamma: \cW(\sS(M),\sigma_M) \to \cW(\sS(\scri),\sigma)$$ determined by the requirement on Weyl generators:
$$\imath_\Gamma(W_M(\phi)) = W_{\scri}(\Gamma \phi)\:.$$
As a consequence of the existence of the $*$-homomorphism $\imath_\Gamma$, one may induce a preferred state 
$\lambda_M$ on the observables in the bulk $M$ form the natural state $\lambda$ defined on the observables 
on the boundary $\scri$. In other words,
{\it the boundary state $\lambda$ can be pulled back to a quasifree state $\lambda_M$ 
acting on observables for the field $\phi$ propagating in the bulk spacetime $M$}: 
\begin{equation}\lambda_M(a):= \lambda(\imath_\Gamma(a))\quad \mbox{for all $a\in \cW(\sS(M), \sigma_M)$} \label{lambdaM} 
\end{equation}
 In \cite{DMP} it has been shown that
 If $(M,g)$ is Minkowski spacetime (so that $(\tM,\tg)$ is Einstein closed universe), hypotheses
{\bf H1} and {\bf H2} are fulfilled so that $\imath_M$ {\em exists}  and {\em $\lambda_M$ coincides with Minkowski vacuum}.
This is not the only case. To illustrate it we recall an important notion.
Let $(M,g)$ be an asymptotically flat vacuum  at future null infinity spacetime. One says that it
admits {\bf future time infinity} $i^+$  if $\exists i^+ \in \tM \cap I^+(M)\:\:\:$ ($i^+ \not \in \scri$) 
such that the geometric extent of $\scri \cup \{i^+\}$ about $i^+$ ``is the same as 
that in a region about the tip $i^+$ of a light cone in a (curved) spacetime''. The rigorous definition has been given by
Friedrich \cite{Friedrich}.

\begin{figure}[th]
\begin{center}
\includegraphics[scale=.7]{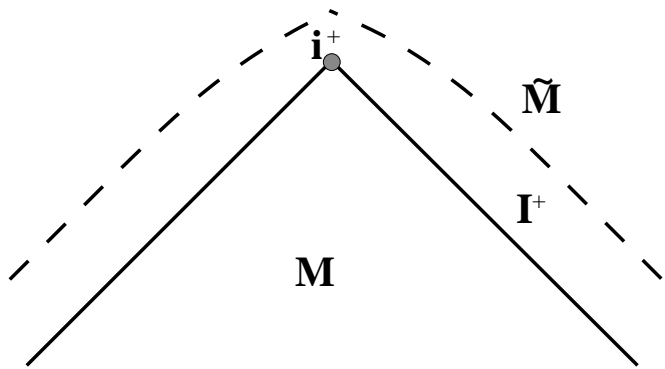}
\end{center}
\caption{Asimptotically flat spacetime with future time infinity $i^+$.}
\end{figure}

We stress that there are lots of Einstein eq.s solutions admitting $i^+$ as established by Friedrich
(actually he considered the past time infinity $i^-$, but the extend is completely symmetric).
With this notion we may state the following result \cite{CMP5} whose proof 
is based on fine estimations of the behaviour about $i^+$ of  $\Gamma \phi$.
The difficult point is to show that $Range [\Gamma] \subset \sS(\scri)$.
Then, barring technicalities, $\sigma(\Gamma \phi_1,\Gamma \phi_2) = \sigma_M (\phi_1,\phi_2)$ 
is a consequence of divergence theorem.\\

\noindent{\bf Theorem 4.} {\it If $(M,g)$ is asymptotically flat at future null infinity, $(M,g)$ and $(\tM,\tg)$ 
are both globally hyperbolic and $(M,g)$ 
admits future time infinity $i^+$, then hypotheses {\bf H1} and {\bf H2} are fulfilled 
so that $\imath_M$ {\bf exists} and 
 $\lambda_M$ {\bf can be induced in the bulk form the $BMS$-invariant state on $\scri$}, 
 $\lambda$.}

\section{Uniqueness of $\lambda$ and remarkable properties of $\lambda_M$.} 
The state $\lambda$ is {\em universal}: it does not depend on the particular bulk
spacetime $M$, but it induces a preferred state $\lambda_M$ on the observables in the bulk 
in any considered asymptotically flat spacetime 
(provided that relevant hypotheses on asymptotic flatness and existence of $i^+$ be fulfilled).
Let us investigate some properties of $\lambda$ and $\lambda_M$ \cite{CMP5,last}.\\

\noindent {\bf Theorem 5}.
{\it Assume that hypotheses (H1) and (H2) hold for the globally hyperbolic a.f. spacetime $(M,g)$ 
(with $(\tM,\tg)$ globally hyperbolic as well)
so that
the state $\lambda_M$ exists.\\
{\bf (a)} $\lambda_M$ is {\bf invariant under} under the (unit component Lie)  group of isometries of $M$, $\{g_t\}$.\\
{\bf (b)} If the Killing generator $\xi$ of $\{g_t\}$ is {\bf timelike and future directed}, then the associated 
unitary $1$-parameter group in the $GNS$ representation of $\lambda_M$ {\bf admits positive self-adjoint generator}
and in the one-particle space there are no zero modes for that generator.}\\

\noindent {\bf Comments on the proofs.}\\
 (a) Let $\{ \beta_{g_t}\}$ and $\{\beta_{g'_t}\}$ be,  respectively,
the  groups of $*$-automorphisms induced by $\{g_t\}$ and $\{g'_t\}\subset G_{BMS}$, acting on bulk 
and boundary observables $a \in \cW(M,\sigma_M)$, $b \in \cW(\scri,\sigma)$
respectively. The action of $\{\beta_{g_t}\}$  on $\lambda_M$ is equivalent to 
 the action of  $\{\beta_{g'_t}\}$ on $\lambda$ and $\lambda$ is $G_{BMS}$-invariant. Therefore:
 $\lambda_M(\beta_{g_t}(a)) = \lambda(\beta_{g'_t}(\Gamma(a))) =
\lambda(\Gamma(a)) = \lambda_M(a)$.\\
(b) If the Killing generator of $\{g_t\}$ is  {\it timelike and future directed} then $\{g'_t\} \subset G_{BMS}$
 is generated by a single {\it causal, future directed} $4$-translation of $T^4$. Then 
 passing to GNS Hilbert spaces, using $\{U_{(\Lambda,\alpha)}\}$, 
one sees, by direct inspection, that $\{U_{\beta_{g'_t}}\}$ has {\it positive self-adjoint generator}.
The analog holds in the GNS representation of $\lambda_M$.\\

Positivity of self-adjoint generators of timelike future-directed $4$-translation of $T^4$
holds true also for $\lambda$. 
It is a remnant of {\bf energy positivity condition} in the bulk. 
Positivity  condition on energy is a stability requirement: it guarantees that, under small (external) perturbations, the system 
does not collapse to lower and lower energy states.
Actually positive energy condition  determines uniquely $\lambda$. 
In fact the subsequent result is valid \cite{CMP5,last}.\\

\noindent {\bf Theorem 6}. 
{\em In the hypotheses of theorem 5,  fix a timelike future-directed 
$4$-translation in $T^4$ and let $\{g_t\}$ be the generated  $1$-parameter $G_{BMS}$-subgroup. Under those hypotheses
the following facts are true.\\
{\bf (a)} If $\omega$ is a pure quasifree algebraic state on $\cW(\sS(\scri), \sigma)$ satisfying both:\\
\indent (i) it is invariant under $\{g_t\}$,\\
\indent (ii)$\{g_t\}$ admits positive self-adjoint generator in the GNS representation of $\omega$,\\
then $\omega$ must be {\bf invariant} under the {\bf whole} BMS group and $\omega$ must {\bf coincide} with $\lambda$.\\
 {\bf (b)}  If $\omega$ is a pure (not necessarily quasifree) algebraic state on $\cW(\sS(\scri), \sigma)$ and it is 
 invariant under $\{g_t\}$, the folium of $\omega$ cannot contain other
 $\{g_t\}$-invariant states.}\\

\noindent {\bf Comments on the proofs.}\\
(a) consequence of  {\em cluster property}: 
$\lim_{t\to +\infty} \omega(a\beta_{g_t}(b)) \to \omega(a)\omega(b)$, $\forall a,b \in \cW(\sS(\scri), \sigma)$
valid for a $\{g_t\}$-invariant pure state, and a uniqueness results  by B.S.Kay
\cite{KayJMP79}. \\
(b) consequence of
of {\em weak asymptotic commutativity} valid for a $\{\beta_{g_t}\}$-invariant pure state $\omega$:\\
$\mbox{w-}\lim_{t\to +\infty}  \left[U(g_t)AU(g_t)^*, B  \right] =0$, 
$\forall A\in \Pi_\omega(\cW(\sS(\scri), \sigma))$,
$B\in \gB(\gH_\omega)$.

\section{The Hadamard property.}  Does $\lambda_M$ satisfy the {\em Hadamard property}? 
if the answer is positive, $\lambda_M$ is a  good starting point for generally covariant and local renormalization procedure, 
in particular it determines a well-behaved renormalized stress-energy tensor
\cite{KW,hadamard}. Let $\omega$ be a regular state on $\cW(M,\sigma_M)$. Let us denote by 
$\omega(x,y)$ the  integral kernel of the two-point function of the state $\omega$: 
\begin{center} $\omega(\Phi(f)\Phi(g)) =  \omega(Ef,Eg)$ \end{center}  
where  $\Phi(f)$ is the standard {\em field-operator} smeared with a test function $f\in C^\infty_0(M) $,
$E = A-R : C^\infty_0(M) \to \cS(M)$  being the
{\em causal propagator} (or ``advanced-minus retarded'' fundamental solution) of Klein-Gordon operator \cite{Wald}. 
The  {\bf global Hadamard property} states that in normal geodesically convex 
neighborhoods of every point of the spacetime:  
\begin{center} $\omega(x,y) = \Delta(x,y)\sigma(x,y)^{-1} + v(x,y) \ln \sigma(x,y) + \mbox{regular function}$ \end{center} 
where $\sigma(x,y)$ is the squared geodesic distance and $\Delta$, $v$ depend on the local geometry only. 
The {\bf global Hadamard property}
is similar, but it involves the (complicated) behaviour of the two-point function in a neighborhood of a Cauchy surface of $M$
\cite{KW}. This is a condition {\em very} difficult to check directly for $\omega = \lambda_M$!\\
Radzikowski found out a micro-local characterization of the (global) Hadamard condition \cite{Rad}: 
{\em If} $\omega \in \cD'(M\times M)$, that is if $\omega$ is a distribution on $M\times M$ thus satisfying continuity
with respect to the relevant seminorm topologies,
the global Hadamard property is {\bf equivalent} to a specific shape of {\em wave front set} of $\omega$, $WF(\omega)$.
More precisely $WF(\omega)$ is made of the elements 
$(x,{\bf p}_x,y,-{\bf p}_y) \in T^*(M\times M) \setminus 0$
such that:\\
 (1) ${\bf p}_x$ is future directed and \\
 (2) there is a null geodesic from $x$ to $y$ having there cotangent vectors 
${\bf p}_x$ and $-{\bf p}_y$ respectively.)\\
Using Radzikowski framework as far as the item (b) has been concerned,  the following final result has been  recently obtained \cite{last}.\\

\noindent {\bf Theorem}. {\em Assume the hypotheses of theorem 5. The following facts are true.\\
{\bf (a)} $\lambda_M \in \cD'(M\times M)$ also if there are 
bad compositions of distributions ($\sS(\scri)$ nonstandard space of test functions).\\
{\bf (b)} $\lambda_M$ is globally Hadamard on $M$.} \\

\noindent {\bf Comments on the proof.}\\
The proof of (b) has been performed establishing first the validity of the local Hadamard condition. Then 
the global Hadamard property has be reached using a ``local-to-global'' argument
introduced by Radzikowski in the second paper in \cite{Rad}. 

\section{Final comments.} 

The unique, positive energy,  $BMS$-invariant, quasifree, pure state $\lambda$ 
is completely defined using the universal structure of the class of  asymptotically flat 
 vacuum spacetimes at null infinity, no reference to any particular spacetime is necessary. 
 In this sense $\lambda$ is {\bf universal}.
On the other hand  $\lambda$ induces a well-behaved quasifree state $\lambda_M$ in each asymptotically flat spacetime 
$M$ admitting $i^+$. 
$\lambda_M$ is quite natural: it coincides with Minkowski vacuum when $M$ is Minkowski spacetime, $\lambda_M$ it is invariant under 
every isometry of $M$ and fulfills the requirement of energy positivity with respect to every 
timelike Killing field in $M$.
$\lambda_M$ may have the natural interpretation of {\bf outgoing scattering vacuum}. 
Finally $\lambda_M$ has been showed  to verify the Hadamard condition and therefore 
it may be used as background 
for perturbative procedures (renormalization in particular), and it provides a natural notion of massless 
particle also in absence of Poincar\'e symmetry but in the presence of asymptotic flatness. 
Indeed  all the construction works for {\em massless} fields (with conformal coupling). 
What about {\em massive} fields?\\
How to connect bulk massive fields to $BMS$-massive fields on $\scri$
and to known unitary representations of $G_{BMS}$ with $m_{BMS} >0$? \cite{AD}.
This is an open issue which deserves future investigation

\end{document}